# Eventful: Crowdsourcing Local News Reporting

ELENA AGAPIE, Microsoft Research
ANDRÉS MONROY-HERNÁNDEZ, Microsoft Research

## 1. INTRODUCTION

We present Eventful, a system for producing news reports of local events using remote and locative crowd workers. The system recruits and guides novice crowd workers as they perform the roles of *field reporter*, *curator*, or *writer*. Field reporters attend the events in person, and use Eventful's mobile web app to get a personalized mission, submit content, and receive feedback. Missions include tasks such as taking a photo, and asking a question to an attendee. In parallel, remote curators approve, reject, and give real-time feedback on the content collected by field reporters. Finally, writers put together a report by mashing up and tweaking the content approved by the curators. We used Eventful to produce a news report for each of the six local events we decided to cover as we piloted the system. The process was typically completed under an hour and costing under $150 USD.

Eventful is motivated by the desire to explore new mechanisms to fulfill the need for local journalism in a time when local news outlets struggle to survive. News organizations have reduced the resources devoted to report local events because revenue has declined [Pew 2010]. This absence of local news presents serious challenges to the health of local communities as it creates a vacuum of accountability for local government and businesses [Patterson 2007].

Professional journalism is a complex endeavor that we are not claiming to replace with Eventful. However, we are inspired by citizen journalism as a model that opens up new possibilities for non-experts to perform journalistic tasks. Citizen journalists have covered important events like terrorist attacks [Cassa et al. 2013] and revolutions [Lotan et al. 2011]. Even large news corporations, such as CNN [iReport] and The Guardian [Witness], have formalized the contributions of citizen journalists by inviting people to provide tips, photographs, and videos.

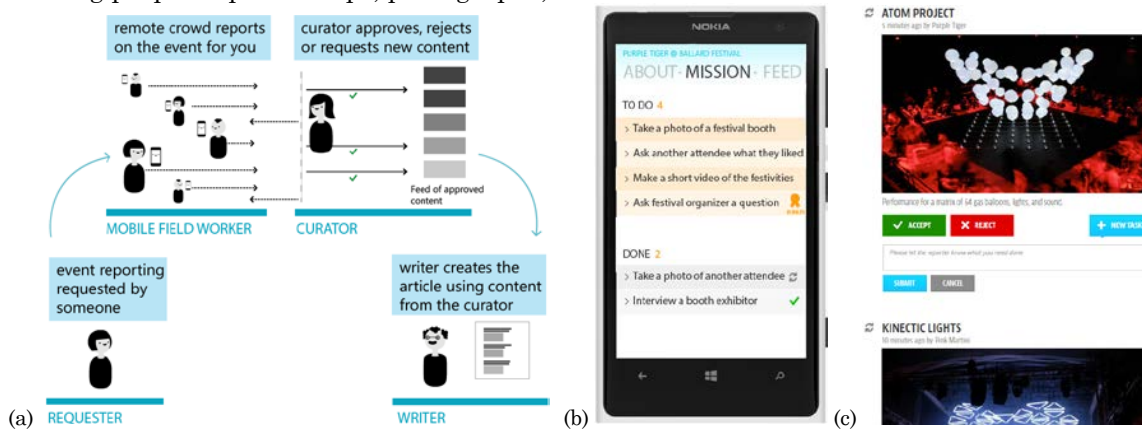

Fig. 1. (a) System flow. (b) Mobile Mission App lists the tasks that workers received, along with their status. (c) Feed where curator accepts, rejects, request tasks.

One of the challenges with citizen journalism is that it often depends on a single individual to produce a news report, such as a blog post. Social media has begun to address this challenge by enabling people to provide small contributions, such as short notes on Twitter, quick photos and videos on Instagram.

However, these contributions tend to be scattered, unstructured, and undirected, and very much dependent on the work of volunteers who happen to be motivated to contribute at the right time and place. For example, large crises events motivate people to contribute information that then can be curated by other volunteers [Starbird 2013]. Similarly, the possibility to be featured by a large news organization might provide the necessary impetus for people to contribute information to requests such as the ones posted by CNN iReport.

Leaving the motivation issues aside, even if individuals want to cover a local event, people don't have easy access to guidelines on how and what information to collect when covering an event. Furthermore, there is no simple way for many contributions to be aggregated into an article. System research on crowdsourced news reporting tools is limited but the work by Väätäjä and colleagues [2012] demonstrated how a system can connect newsroom members with citizens by asking people to submit specific content that is then used in the professional news outlet. However, their system was reliant on journalists to select the right content, did not focus on time-constrained events, and did not allow for real-time feedback to volunteers playing the role of field reporters. More broadly, crowdsourcing research has previously shown writing journalistic articles is possible with the help of novice crowds [Bernstein et al. 2010, Kittur et al. 2011]. We aim to build on this for the production of news reports.

In this research, we focus on two particular challenges: a) working with locative crowdworkers to report on events that are too small to receive mainstream media attention, and b) offering guidance to workers, a priori and during the event. We will describe the tool we built to automate parts of the process and a series of event reports used as case studies for our current design.

## 2. EVENTFUL: SYSTEM DESCRIPTION

As the system evolved, more parts of the processes were handled by the system itself, while at the beginning a number of steps were performed manually by the researchers. The system aims to enable people, such as community members, local bloggers, event organizers, or even journalists, to request a report for an event. The requester enters the name, location, date, time, and type of the event (e.g., town hall meeting, festival, sports game, etc.). These characteristics determine the set of missions associated with the event. Combinations of tasks are packaged into *missions*. Each missions set is assembled into a *template*. Templates can be reused for events with similar characteristics. The whole process is intended to be finished in under an hour and cost under a hundred dollars. System components:

1. **Mission templates.** Each mission assigned to a crowd worker contains a list of tasks. The system has mission templates for each event type. For example, one set of missions for a town hall meeting, and a different set for a festival. We designed the templates and assessed the type of template to be chosen for each event.
2. **Mission App.** This is *a mobile web app that shows field reporters their assigned mission as a list of tasks they need to complete. The app facilitates the submission of content over email. Tasks have two states: "to do" and "done". A remote curator can approve tasks in real time, offer feedback (e.g., "next time focus on this specific aspect of a photo"), and create bonus tasks (Fig. 1.b).*
3. **Live Feed.** This page shows the stream of content and allows curators to accept or reject submissions (Fig. 1.c). Information is organized into predefined article sections (e.g., the audience impressions of an event, photos, and interviews).

The system recruits workers from TaskRabbit.com, a locative crowdsourcing platform, gives them missions, guides them as they perform their tasks, and provides them feedback in real-time. Field workers produce photos, video or audio recordings, interviews or text updates. The content is aggregated in a feed that gets curated by another group of crowd workers. Finally, a third set of crowd workers composes a report or an article.

3.  PILOT STUDIES

We covered six events that were unlikely to be covered by the mainstream media (Table 1). The town hall meeting (2) was selected in collaboration with a well-known neighborhood blogger who then featured the report on her blog [Rainier]. For each event, the system generated a report usually within an hour of the end of the event. As the events progressed, more and more of the organizational part of process was handled automatically by the system. The last event had 24 unpaid volunteers who acted as field reporters and one paid worker who produced the report. The other events got covered by paid workers.

| Event | Cost | Workers | Report Length | Submitted Items |
|---|---|---|---|---|
| Festival | $122 | 2 | 382 words | 27 |
| Art Show | $107 | 2 | 438 words | 32 |
| Public Talk | $147 | 4 | 363 words | 50 |
| Town hall (1) | $98 | 4 | 180 words | 10 |
| Town hall (2) | $60 | 4 | 183 words | 23 |
| Hackathon | $31 | 25 | 296 words | 83 |

Table 1. Details of the event covered, workers, and output.

In the early events, we spent quite a bit of time explaining tasks, and supervising workers. We minimized this by first hiring a person to manage the workers, and then by offloading most of this coordination work to the system, resulting in a clearer workflow. That said, we encountered two main challenges that are worth further investigation:

1. **Existing vs new tools**. Well-known tools (e.g., email or Twitter) for field reporters to submit content and receive feedback was advantageous because they did not have to learn something new. However, no tools were universally well-known, and none of them were structured enough for missions. The "Mission App" was an attempt at structuring this at the expense of having people to learn yet another tool. The app still relied on email as the channel for content submission, which came with the challenge that some people had never used their email on their phone before.
2. **Community members vs paid workers.** Overall, paid workers tended to deliver more content than members of the community who could volunteer to perform the missions. This could be partly explained by the possibility that for some people the high cognitive load needed to perform the missions detracts them from enjoying the event. The disadvantage of paid workers was, however, that they lacked inside knowledge that would infuse the reports with the kind of "bias" needed by the report's audience, which was something the "town hall meeting" blogger mentioned.

4.  CONCLUSION

We presented a vision for a system that provides "journalism as a service" using locative and remote crowdsourcing. We believe that given the right workflows and technologies, a system like Eventful could enable a sustainable end-to-end solution for local news production for the people and by the people. Such system would require an "interest aggregation" and a "resource mobilization" component, akin to crowdfunding platforms, that aggregates a community's financial, and time-commitment contributions for completing the necessary missions to cover issues of interest to such community. We hope this work inspires future research to investigate new forms of journalism that rely on both paid workers and volunteers.

5.  ACKNOWLEDGEMENTS

Thanks to FUSE Labs at MSR for supporting this work, to Todd Newman, Janice Von Itter, Melissa Quintanilha for the software development, interface design, respectively.